\PassOptionsToPackage{hyphens}{url}
\documentclass[letterpaper,twocolumn,10pt]{article}
\usepackage{usenix_2020_09}
\usepackage{epsfig,endnotes}
\usepackage{paralist}
\usepackage{enumerate}
\usepackage{graphicx}
\usepackage{color}
\usepackage{url}
\usepackage{booktabs}
\usepackage{amsmath}
\usepackage{amssymb}
\usepackage{mathtools}
\usepackage{multicol}
\usepackage{balance}
\usepackage{caption}
\usepackage{bigstrut}
\usepackage{xcolor,colortbl}
\usepackage{mathtools, nccmath}
\usepackage{listings}
\usepackage{hyperref}
\usepackage[utf8]{inputenc}
\usepackage{algorithm,algpseudocode}
\usepackage{enumitem}

\newcommand{\Web}{Web}
\newcommand{\EG}{e.g.}
\newcommand{\IE}{i.e.}
\newcommand{\JS}{JavaScript}
\newcommand{\EtAl}{et al.}

\newcommand{\NIK}{NetworkIsolationKey}

\newcommand{\TechniqueName}{``pool-party''}
\newcommand{\uTechniqueName}{``Pool-party''}
\newcommand{\Sender}{\emph{sender}}

\newcommand{\Receiver}[1]{\emph{receiver#1}}

\newcommand{\LBU}{limited-but-unpartitioned}

\newcommand{\SSE}{Server-Sent Event}
\newcommand{\NoneSymbol}{$\times$}
\newcommand{\SomeSymbol}{$\ominus$}
\newcommand{\AllSymbol}{$\oplus$}
\newcommand{\SubPoint}[1]{\smallskip\noindent\textbf{#1.} }

\newcommand{\SourceLink}{https://anonymous.4open.science/r/poolparty-2E6B/static/inner.js}

\newcommand{\ChromePlatformDate}{August 9, 2022}

\begin{document}
\sloppy

\title{Pool-Party: Exploiting Browser Resource Pools for Web Tracking}

%\author{}
\author{
{\rm Peter Snyder}\\
Brave Software 
%\\pes@brave.com
\and
{\rm Soroush Karami}\\
University of Illinois at Chicago
%\\skaram5@uic.edu
\and
{\rm Arthur Edelstein}\\
Brave Software 
%\\aedelstein@brave.com
\and
{\rm Benjamin Livshits}\\
Imperial College London 
%\\b.livshits@imperial.ac.uk
\and
{\rm Hamed Haddadi}\\
Brave Software, \/ Imperial College London 
%\\%h.haddadi@imperial.ac.uk
} % end author

\maketitle

\begin{abstract}

We identify class of covert channels in browsers that are not mitigated by 
current defenses, which we call ``pool-party'' attacks. 
Pool-party attacks allow sites to create covert channels by manipulating limited-but-unpartitioned resource pools. 
This class of attacks have been known to exist; in this work we show that they
are more prevalent, more practical for exploitation, and allow exploitation
in more ways, than previously identified. 
These covert channels have sufficient bandwidth to pass cookies and identifiers across site boundaries under practical and real-world conditions.
We identify \emph{pool-party} attacks in all popular browsers, and show they are practical cross-site tracking techniques (i.e., attacks take 0.6s in Chrome and Edge, and 7s in Firefox and Tor Browser).

In this paper we make the following contributions: first, we describe \emph{pool-party} covert channel attacks that exploit limits in application-layer resource pools in browsers. 
Second, we demonstrate that \emph{pool-party} attacks are practical, and can be used to track users in all popular browsers; we also share open source implementations of the attack. 
Third, we show that in Gecko based-browsers (including the Tor Browser) \emph{pool-party} attacks can also be used for \emph{cross-profile} tracking (e.g., linking user behavior across normal and private browsing sessions). 
Finally, we discuss possible defenses.

\end{abstract}

\section{Introduction}
\label{sec:intro}

Browser vendors are increasingly developing and deploying new features to
protect privacy on the \Web{}.  These new privacy features address the most
common ways users are tracked on the \Web{}: partitioning DOM storage to prevent
tracking from third-party state, randomization or entropy reduction to combat
browser fingerprinting, network state partitioning to prevent cache-based
tracking, etc.

However, research has documented other ways \Web{} users can be tracked,
though in ways that may be difficult to conduct under realistic browsing
conditions. Significantly among these are covert-channels that can be
constructed through timing signals, or other side channels. 
These covert-channels allow sites to communicate with each other---or even
other applications---in ways not intended by browsers. 
Such covert-channels can be used to reintroduce the kinds of
cross-site tracking attacks the above-discussed browser protections were
designed to prevent.

Browser vendors have responded to covert-channels in a variety of ways.
Some covert-channels (\EG{} timing signals from abusing HTTP cache state) have
been addressed through platform wide improvements like network state partitioning. 
Other covert-channels have been addressed---or at least mitigated---through
other protections, like isolating sites in their own OS processes. 
Others attacks have been left unaddressed, because browser vendors judge them
to be impractical to execute in realistic browsing scenarios.

In this work we demonstrate that current browser protections are insufficient
to prevent sites from using covert-channels to circumvent anti-tracking
protections in browsers, including the protections deployed by the most
privacy-focused browsers.
We demonstrate this by defining a new category of techniques for
constructing covert-channels, by exploiting the state of
\LBU{} resource pools in the browser. 
Because such covert-channels
are exploited by two parties colluding in the same resource pool,
we call this category of covert-channel \TechniqueName{} attacks.

\uTechniqueName{} attacks create covert-channels out of
browser-imposed limits on pools of resources. When resource pools are
limited (\IE{} the browser only allows pages to access resources up to
some hard limit, after which requests for more resources fail)
and unpartitioned (\IE{} different sites consume resources from a shared pool),
sites can consume and release resources to leak information across
security boundaries. Examples of such boundaries include site boundaries
(\EG{} the browser intends to prevent site A from communicating directly
with site B) and profile boundaries (\EG{} the browser intends sites visited
in a ``standard'' browsing session to not be able to learn about sites
visited in an ``incognito'' browsing session).
More generally, attackers can use these covert-channels to conduct the kinds of
cross-site tracking that the recent browser features were intended to prevent.

We identify practical \TechniqueName{} attacks in all popular browsers, both
in browsers' default configurations, and non-standard, hardened configurations. 
We demonstrate \TechniqueName{} attacks through three resource pools:
WebSockets, \SSE{s}, and Web Workers, and find that all browsers were vulnerable
to at least one form of attack. We further identify other \LBU{} resource pools
in browsers that could be leveraged for \TechniqueName{} attacks. 
Examples of such pools include certain kinds of resource handle
(\EG{} Web Speech API), or limits on how many network requests
(distinct from network connections) can be in flight at once
(\EG{} DNS resolution), among others. 
Finally, we demonstrate that \TechniqueName{} attacks are not just
\emph{theoretical} threats to user privacy, but \emph{practical} threats that
can be used to track users across sites. We show that in Gecko-based browsers
(including the Tor Browser), \TechniqueName{} attacks can create
covert-channels \emph{across profiles}, allowing sites to link behaviors in
``private browsing'' modes with standard, long term browser identities.

These findings are important for the development of browser partitioning.
All browser engines support some forms of partitioning:
WebKit partitions DOM storage and some kinds of network state, Gecko partitions
DOM storage and network state, and Chromium partitions
network state\footnote{At time of writing, network state partitioning is deployed for a portion of Chrome and Edge users, as part of the ``NetworkIsolationKey'' feature}. 
Brave has extended Chromium to also partition DOM storage.

\subsection{Contributions}

This work makes the following contributions:

\begin{compactitem}
    \item We \textbf{define a new category of technique for creating covert-channels in browsers}.
        We call this category of covert-channel \TechniqueName{} attacks, and describe how
        the approach differs from the kinds of privacy attacks browsers currently aim to defend against;
    \item We \textbf{evaluate deployed browsers}, and find all popular browsers
        and browser engines are vulnerable \TechniqueName{} attacks;
    \item We provide three \textbf{open-source, proof-of-concept} implementations of
        our attack that work in all browsers\footnote{\SourceLink{}};
    \item We perform a \textbf{performance measurements} to evaluate the bandwidth
        and practicality of \TechniqueName{} attacks, and find that
        \TechniqueName{} attacks are a practical basis carrying out
        cross-site tracking attacks;
    \item We \textbf{discuss potential mitigation strategies} for how browsers
        could defend against \TechniqueName{} attacks.
\end{compactitem}

\subsection{Responsible Disclosure}
We have presented our findings to the following browser vendors
(in alphabetical order): Apple, Brave, Google, Microsoft, Mozilla, Opera,
Tor Project.  All reports were made over 90 days in advance of this submission.

Microsoft and Opera responded that since the discussed vulnerabilities were in
Chromium, they would wait for Google to address the problem. The Tor Project
similarly said they would rely on Mozilla to address the
vulnerabilities\footnote{\url{https://gitlab.torproject.org/tpo/applications/tor-browser/-/issues/41381}}.

Some vendors have shipped fixes for the vulnerabilities identified in this work.
Safari fixed the SSE event vulnerability in version
15.2\footnote{\url{https://support.apple.com/en-us/HT212982}}, and Brave has
released fixes for the WebSocket\footnote{\url{https://github.com/brave/brave-core/pull/11609}}
and SSE\footnote{\url{https://github.com/brave/brave-core/pull/16882}} vulnerabilities.

Google\footnote{\url{https://bugs.chromium.org/p/chromium/issues/detail?id=1249658}}
and Mozilla\footnote{\url{https://bugzilla.mozilla.org/show_bug.cgi?id=1730797}} also
plan to address these vulnerabilities, though have not done so yet.
These organizations are focusing on a mixture of browser-wide fixes
(\IE{} comprehensively partitioning all resource pools, not only the
resource pools discussed in this work) and updates to Web standards (\EG{}
defining limits and the scope of connection pools).

\section{\uTechniqueName{}: Definition and Background}
\label{sec:definitions}

This section provides context for how \TechniqueName{} attacks
relate to other \Web{} tracking techniques, and how browser vendors'
privacy models and goals have changed.
% Unfettered communication across 
% site boundaries used to be an acceptable (if not intended) behavior in many browsers.
% Now all browsers include modes that try to prevent privacy-harming cross-site
% communication (though these modes are not enabled by default in all browsers).
% \TechniqueName{} attacks allow trackers to circumvent these privacy-protections
% in all popular browsers.
This section also describes \emph{why} existing browser protections
fail to protect users against \TechniqueName{} attacks.

This section first defines ``\Web{} tracking'', followed by discussing
how privacy models in \Web{} browsers have improved, and why \TechniqueName{} attacks allow
trackers to violate intended privacy boundaries in all popular browsers.
Next, we describe how \TechniqueName{} attacks relate to both i) other covert-channels
in browsers, and ii) conventional \Web{} tracking techniques.  We then explain why 
existing browser protections do not protect users against
\TechniqueName{} attacks, and conclude by describing how
this work relates to a category of attack previously known to be possible, but not
thought to be practical.

\subsection{Web Tracking and Cross-Site Tracking}
\label{sec:definitions:tracking}
This sub-section gives a working definition of \Web{} tracking. Our goal
is not to provide a formal, unambiguous definition (the
phrase ``\Web{} tracking'' is used too broadly to likely allow for one),
but instead to give a practical definition to build
on through the rest of this work.

We use ``\Web{} tracking'' to refer to a user
being re-identified across conceptual contexts, without the
user's expectation or consent. We use ``context'' to refer to
a grouping of activities that the user expects to be separate from, and not accessible to,
other similar contexts. This definition is similar to the W3C's proposed
privacy principals\footnote{\url{https://w3ctag.github.io/privacy-principles/}}.

Contexts might be divided by \emph{time} (\EG{} a site
re-identifying a user revisiting the same site a week after first visiting,
despite the user clearing browsing data), \emph{application} (\EG{} a
site re-identifying a user visiting a site in Safari as the same user
who previously visited the site in Chrome), \emph{profile} (\EG{} a
site identifying that the visiting in an private/incognito browser session
is the same user visiting the site in a standard browser session), or 
\emph{site} (\EG{} two sites colluding to learn that browser sessions
occurring on each site belong to the person). The commonality
is the users' reasonable expectation that things that happen in one context
are not readily known and available to other contexts.

\subsection{First-Party Site as Privacy Boundary}
Browser vendors are converging on the first-party site
as the Web's privacy boundary. All browsers include features intended to prevent
sites from communicating across first-party site boundaries. Some
browsers enforce this boundary by default; others only do so
with opt-in ``privacy'' modes, but all browsers include such features.
% The significance of this work is that we show that \TechniqueName{} techniques
% allow for practical and real world violations of this privacy boundary in all popular browsers.

% Broadly, browsers either use, or include the option to use, the first-party site (as determined by the effective-top level
% domain, \texttt{eTLD+1}, of the top level document) as a privacy boundary.
% Scripts, resources, and sub-documents embedded in different pages under a first-party
% share identifiers and are allowed to communicate with each other.
% However, scripts and resources are not intend to be able to (re)identify a user
% \emph{across} sites. This privacy boundary is intended to hold
% both when browser features are used in their intended ways (\EG{} cookies, DOM storage),
% and when such features are abused (\EG{} abusing cache or network state).

% We provide more detail about how browser features are (ab)used for cross-site tracking
% in Section \ref{sec:definitions:contrasts}.

Using the first-party site as a privacy boundary means that
a third-party embedded under two different first-parties
should not be able to confidently know it was the same person visiting each site,
unless the user intentionally re-identifies themselves to the third-party.

The rest of this subsection documents that, and in what configurations, each
browser uses the first-party site as their privacy boundary. In all the
discussed configurations, browsers intended to communication
across first-party site boundaries, and in all cases attackers
can circumvent the intended privacy boundary through \TechniqueName{} attacks.

\SubPoint{Gecko Browsers}
Both Firefox (as of version 103) and Tor Browser enforce the first-party site as the privacy boundary by default.
In Tor Browser, the protection is sometimes called ``first-party isolation'' or ``cross-origin identifier
unlinkability''\footnote{\url{https://2019.www.torproject.org/projects/torbrowser/design/##identifier-linkability}}.
In Firefox, the feature is called ``Total Cookie
Protection''\footnote{\url{https://blog.mozilla.org/security/2021/02/23/total-cookie-protection/}}.

\SubPoint{WebKit Browsers}
Safari uses the first-party site as the privacy boundary by default. The WebKit
documentation makes this privacy boundary explicit in their documentation,
which mentions that they intended to protect against cross-site communication
through covert-channels\footnote{\url{https://webkit.org/tracking-prevention-policy/}}.

\SubPoint{Chromium Browsers}
Chromium \emph{does not} enforce the first-party site as a privacy
boundary by default. However, Chromium allows for configurations that do, by a combination
of i) disabling third-party cookies (to prevent DOM storage communication
across site boundaries) and ii) enabling Chromium's \texttt{``\NIK{}''}
(NIK)\footnote{\url{https://github.com/shivanigithub/http-cache-partitioning\#choosing-the-partitioning-key}}
features (which partition caches and other network state by first-party).

Neither Chrome or Edge disable third-party storage by default, but both do enable
NIK features for most users.
We note though that even when Chrome and Edge are configured to use the first-party site
as the privacy boundary, those browsers are vulnerable to \TechniqueName{} attacks.

The Brave Browser uses a modified version of Chromium that, by default uses
the first-party site as a privacy boundary.  It does this by partitioning third-party DOM
storage by first-party\footnote{\url{https://brave.com/privacy-updates/7-ephemeral-storage/}},
and by enabling (and extending) Chromium's NIK system for all users\footnote{\url{https://brave.com/privacy-updates/14-partitioning-network-state/}}.

\subsection{Description of \uTechniqueName{} Attack}
\label{sec:definitions:pp}

\uTechniqueName{} attacks manipulating pools of browser resources
which are limited (\IE{} the browser restricts how many of the resource 
can be used at one time) and unpartitioned (\IE{} different contexts consume
resources from the same pool). While the examples focused on in this work
utilize either \LBU{} pools of i) network connections or ii) thread handles, browsers include
many other \LBU{} resource pools that could be similarly exploited, such as
pools of file handles, subprocesses, or other resource handles.

A \TechniqueName{} attack occurs when parties operating in distinct contexts
(contexts the user expects to be distinct and blinded from each other)
intentionally consume and query the availability of the limited resources in
a resource pool, to create a cross-context communication channel. Each context
can then use the communication channel to pass an identifier, allowing each
party to link the user's behavior across the two contexts. We note again that
most commonly the two contexts considered here are two different websites running
in the same browser profile, but could also be the same (or different) websites
running in different browser profiles.

\begin{algorithm}[t]
  \small
  \begin{algorithmic}
    \State \text{\textbf{Site A:} }$I_a \gets random\text{ }N\text{ }bits$
    \State \text{\textbf{Site B:} }$I_b \gets \text{\textit{empty string}}$
    \While{$i \gets I_a$}
      \State \text{\textbf{Site A:} }Stop any playing videos
      \If{$I_a[i] = 1$}
        \State \text{\textbf{Site A:} }Play a video
      \EndIf
      \State Wait 5 seconds
      \If{Site B is able to play a video}
        \State \text{\textbf{Site B:} }$I_b[i] = 1$
      \Else
        \State \text{\textbf{Site B:} }$I_b[i] = 0$
      \EndIf
    \EndWhile
  \end{algorithmic}
  \caption{Toy example of a \TechniqueName{} attack.}
  \label{algo:toy-algo}
\end{algorithm}

Algorithm \ref{algo:toy-algo} presents a simple-though-limited technique for
conducting a \TechniqueName{} attack, where sites can trivially transform this
optimization choice into a cross-site tracking mechanism. 

For this toy example, assume a browser vendor wants to improve performance
by only allowing one video element to be loaded at a time, across all sites.
If a video is currently playing on any page, the site will receive an error
if it tries to play a new video. An attacker use this implementation
choice to ``send'' a bite across site-boundaries by playing (or not)
a video on one site, and checking on another site whether there is a video playing.
An arbitrarily large message can be sent by repeating this process.
A more realistic and efficient
technique is presented in Section~\ref{sec:algo}.

\subsection{Relationship to Other Covert-Channels}
\label{sec:definitions:covert-channels}
\uTechniqueName{} attacks differ from other covert-channel attacks by targeting intentional, application-imposed limits. This differs from many other
covert-channel attacks in two ways, both of which increase the
practicality of \TechniqueName{} attacks.

First, \TechniqueName{} attacks
target application-level resources, while many other covert-channels 
target parts of the system below, or at least distinct from, the
application (\EG{} hardware restrictions like CPU caches,
OS details like interrupt schedules or memory management, or
language runtime features like garbage collection). This is significant because,
the lower in the stack the attacker targets, the more likely the resource is (all other things
being equal) to be shared with other actors on the system.
This means that lower-level covert-channels are more likely to be noisy, and so more difficult
to communicate over.

Second, related but distinct, \TechniqueName{} attacks target browser-managed
resources, resources that are, in most cases, intentionally shielded from
other applications on the system. This again reduces the chance that
colluding parties will have to contend with a noisy, unpredictable covert-channel.

\subsection{Relationship to Other Tracking Methods}
\label{sec:definitions:contrasts}
\uTechniqueName{} attacks do not fall neatly into the categories usually used
to describe browser tracking techniques. This subsection briefly describes
the rough-taxonomy used in online-tracking research, and why \TechniqueName{}
does not cleanly fall into existing categories.

\SubPoint{Stateful Tracking}
\label{sec:definitions:contrasts:stateful}
``Stateful tracking'' most commonly refers to websites using explicit storage APIs
in the Web API (\EG{} cookies, localStorage, indexedDB) to assign identifiers
to browser users, and then read those identifiers back in a different context,
to link the identity (or, browser behavior) across those contexts.

Stateful-tracking also describes other ways websites can set and
read identifiers, by using APIs and browser capabilities not intended for
such purposes. Examples of such techniques include exploiting the browser
HTTP cache, DNS cache or other ways of setting long term state (\EG{} HSTS
instructions~\cite{syverson2018hsts}, favicon caches~\cite{favicon2021}, or,
ironically, storage intended to prevent tracking~\cite{janc2020information}).

\begin{table}[t]
  \begin{center}
    \small
    \begin{tabular}{lll}
      \toprule
       \bf  Browser & \bf DOM Storage   & \bf Network State   \bigstrut[t]\\
      \midrule
        Brave   & \AllSymbol{}  & \AllSymbol{} \\
        Chrome  & \NoneSymbol{} & \SomeSymbol{} \\
        Edge    & \NoneSymbol{} & \SomeSymbol{} \\
        Firefox & \AllSymbol{}  & \AllSymbol{} \\
        Safari  & \AllSymbol{}  & \AllSymbol{} \\
        Tor Browser & \AllSymbol{}  & \AllSymbol{} \\
      \bottomrule
    \end{tabular}
  \end{center}
  \caption{State partitioning features in popular browsers (in alphabetical
    order). \AllSymbol{}, \SomeSymbol{} and \NoneSymbol{} indicate the
    feature being available by default for all, some, or no users, respectively.}
  \label{table:browser-partitioning}
\end{table}

Browsers increasingly protect users from stateful tracking by
partitioning storage by context, mostly commonly be the effective-top level
domain (\IE{} \texttt{eTLD+1}) of the website. Giving each context a unique
storage area prevents trackers from reading the same identifier across multiple
contexts, and so prevents the tracker from linking browsing behaviors in
different contexts. Partitioning explicit storage APIs is often referred to
as \textbf{DOM Storage} partitioning.  Partitioning caches and other ``incidental''
ways sites can store values is often called \textbf{network state partitioning}.
Table \ref{table:browser-partitioning} provides a summary of state partitioning
in popular browsers.

Browser state partitioning strategies fail to defend against \TechniqueName{}
attacks because \TechniqueName{} attacks do not rely on setting or retrieving
browser state (at least not in the way state is generally discussed in this
context, meaning the ways that sites can write state to the users profile).
\uTechniqueName{} attacks instead rely on implementation details of browser
architecture, where device resources are \LBU{}. While partitioning strategies
could also be used to defend against \TechniqueName{} attacks, as discussed in
Section \ref{sec:discussion}, certain aspects of these attacks make partitioning
approaches difficult in practice.

\SubPoint{Stateless Tracking}
\label{sec:definitions:contrasts:stateless}
``Stateless tracking,'' (also often called ``browser fingerprinting'') refers to
the category of \Web{} tracking techniques whereby the attacker constructs a
unique identifier for the user by combining a large number of
semi-distinguishing browser and environmental attributes into a stable, unique identifier. Examples of such semi-distinguishing features include the operating system the browser is running on, the browser version, the names and the number of plugins or hardware devices available, and the details around the graphics
and audio hardware present, among many others~\cite{laperdrix2020browser}.

In contrast to ``stateful'' tracking techniques, ``stateless''
techniques do not require sites to be able to set and read identifiers across
context boundaries, and so are robust to storage partitioning defenses.
Stateless attacks instead rely on three conditions to be successful:

\begin{compactitem}
  \item \textbf{Coordination:} code running in different contexts must know
    to query the same (or at least sufficiently large intersection of) browser
    attributes.
  \item \textbf{Stability:} the browser must present the same values for the
    same semi-distinguishing attributes across contexts (otherwise the browser
    will yield different fingerprints in different contexts, preventing
    the attacker from matching the two fingerprints).
  \item \textbf{Uniqueness:} the browser must present enough semi-distinguishing attributes to allow the site to accurately
    differentiate between users (otherwise the attack will confuse two
    different users as the same person)
\end{compactitem}

\begin{table}[t]
  \begin{center}
    \small
    \begin{tabular}{lccc}
      \toprule
        \bf Browser & \bf Coordination   & \bf Stability  & \bf Uniqueness   \\
      \midrule
        Brave   & \AllSymbol{}  & \AllSymbol{}  & \AllSymbol{} \\
        Chrome  & \NoneSymbol{} & \NoneSymbol{} & \NoneSymbol{} \\
        Edge    & \NoneSymbol{} & \NoneSymbol{} & \NoneSymbol{} \\
        Firefox & \AllSymbol{}  & \NoneSymbol{} & \SomeSymbol{} \\
        Safari  & \NoneSymbol{}  & \NoneSymbol{} & \AllSymbol{} \\
        Tor Browser & \NoneSymbol{} & \NoneSymbol{} & \AllSymbol{} \\
      \bottomrule
    \end{tabular}
  \end{center}
  \caption{Stateless tracking protections in popular browsers (in alphabetical
    order). \AllSymbol{}, \SomeSymbol{} and \NoneSymbol{} indicate the
    defense is available by default, off by default, or not
    available, respectively.}
  \label{table:fingerprinting-defenses}
\end{table}

\noindent
Browsers defend against ``stateless'' trackers by attacking any of these
three requirements. A browser might prevent \textbf{coordination} by blocking
fingerprinting code on sites, or prevent \textbf{stability} by making the
browser present different attributes to different sites (such as in
\cite{laperdrix2017fprandom, nikiforakis2015privaricator}), or prevent
\textbf{uniqueness} by reducing the entropy provided by each attribute. Table
\ref{table:fingerprinting-defenses} provides a summary of deployed
``stateless'' defenses in popular browsers.

Browser defenses against ``stateless'' tracking techniques fail to defend
against \TechniqueName{} attacks because of differences in the nature of the
attack. ``Stateless'' techniques target stable semi-distinguishing browser
characteristics which are set by a page's execution environment.
\uTechniqueName{} attacks, in contrast, are enabled by sites consuming
and reading the availability of limited resources in the browser across
execution contexts. Therefore, unsurprisingly, browser defenses against
``stateless'' tracking provide no protection against \TechniqueName{} attacks.

\SubPoint{XS (Cross-Site) Leaks}
\uTechniqueName{} attacks are most similar to a category of attack loosely
called ``XSLeaks''\footnote{\url{https://xsleaks.dev/}}, a broad collection of
ways sites can send signals to each other in ways generally unintended by
browser vendors. However, we note that in contrast to ``stateful'',
``stateless,'' \TechniqueName{} attacks, XSLeaks do not have a common cause
or remedy; instead, XSLeaks can be largely thought of as a catchall
for cross-site (or cross-context) techniques that do not not fit in another
category. Examples of XSLeaks include timing channels stemming from a variety of
causes, unintended side effects of experimental browser
features\footnote{\EG{} \url{https://xsleaks.dev/docs/attacks/experiments/scroll-to-text-fragment/}},
or misuse of other browser APIs\footnote{\EG{} \url{https://xsleaks.dev/docs/attacks/window-references/}}.

The lack of a common cause of XSLeaks makes it impossible to generalize about
defensive strategies or deployed browser defenses. Recent work in this area has 
identified ways sites can leak information across browser-imposed boundaries,
including through unintended side effects in how browsers handle errors, implement
cross-origin opener-policy (COOP), cross-origin resource policy (CORP),
and cross-origin read blocking (CORB) policies,
or limit the length of redirection chains, among many other signals~\cite{knittel2021xsinator}.

We note though that \TechniqueName{} attacks are most common to the ``connection pool'' attacks identified by the XSLeaks
project\footnote{\url{https://xsleaks.dev/docs/attacks/timing-attacks/connection-pool/}}.
This work makes the following contributions beyond the issues documented
by the XSLeaks project, and the related work done by Kinttel \EtAl{}~\cite{knittel2021xsinator}.

\begin{enumerate}
    \item This work defines a larger category of attack than XSLeaks,
    where any \LBU{} resource pool can be transformed into a covert-channel.
    The attack documented by the XSLeaks project is 
    a subset of the larger category of attack discussed in this work.
    Network connection pools can be abused to conduct \TechniqueName{} attacks,
    but other kinds of resource pool can too.
    For example, the Web Workers pool in Firefox thecan be exploited to
    conduct \TechniqueName{} attacks. Section~\ref{sec:discussion:additional}
    identifies additionalm non-network-connection pools that can be exploited.

    \item We demonstrate that attacks of this type are not just theoretically
    possible, but are practical, and are real-world threats to \Web{} privacy.

    \item The ``connection pool'' attack identified by the XSLeaks project relies on abusing the connection pool to create timing channels, while \TechniqueName{} attacks utilize the number of available resources in the pool to create the
    communication channel. This small difference is significant. Using
    the amount of available of resources in the pool as the communication channel
    makes the attack more robust to noise introduced from other sites, and
    mitigations against one form of the attack may not apply to the other.
\end{enumerate}
\section{Generic \uTechniqueName{} Attack Algorithm}
\label{sec:algo}

In the previous section we presented the category of \TechniqueName{} attack in the
abstract, explained why \TechniqueName{} attacks are different from other
attacks discussed previously in the literature, and why current browser defenses fail to
protect against \TechniqueName{} attacks. In this section we present a generic
algorithm for conducting \TechniqueName{} attacks, which can then be applied
to any resource pool in a browser where the following conditions are met.

\begin{enumerate}
  \item The resource pool is \textbf{limited}, meaning that sites can request resources
    from the pool until a global limit is hit, after which sites are prevented from accessing more
    resources, in a manner the site can detect.
  \item The resource pool is \textbf{unpartitioned}, meaning that different
    contexts (\EG{} sites, profiles, etc.) all draw from the same global resource
    pool. Put differently, the attack will fail if each context gets a
    distinct resource pool.
  % This item seems redundant with the first one --> \item Sites can \textbf{probe} the resource pool, and learn whether the pool has been exhausted, or if some resources remain.
  \item Sites can \textbf{consume} resources from the pool without restrictions, as long as the pool is not already exhausted.
    \item After consuming resources, sites can \textbf{release} any number of those resources back into the pool.
\end{enumerate}
Any resource pool where the above four criteria are met can be transformed
into a covert communication channel between any two parties sharing the
resource pool.

We have identified resource pools matching the above criteria in current versions
of all popular browsers, even browsers that particularly emphasize their privacy features
(\EG{} Brave Browser, Tor Browser), and even when browsers are ``hardened'' by the
enabling of non-default, privacy-focused features (as discussed in
Section~\ref{sec:definitions:contrasts:stateful}).

\subsection{\uTechniqueName{} Algorithm}
\label{sec:algo:desc}

\begin{algorithm}[ht!]
  \small
  \raggedright
  \begin{algorithmic}
    \State \text{\underline{\textbf{Inputs.}}}
    \State $\text{POOL\_SIZE} \gets \text{size of resource pool}$
    \State $\text{PKT\_SIZE} \gets \lfloor\log(POOL\_SIZE)\rfloor$
    \State $\text{MSG} \gets \text{binary string to transmit}$
    \State $\text{NEGOTIATE\_INTERVAL} \gets \text{time to choose sender and receiver roles}$
    \State $\text{PULSE\_INTERVAL} \gets \text{time to transmit one chunk of data}$
    \State \text{ }

    \State \text{\underline{\textbf{1. Setup.}}}
    \State $\text{CHUNKS} \gets \text{MSG split into packets of size PKT\_SIZE}$
    \State $\text{RECV\_MSG} \gets \text{\textit{empty string}}$
    \State $\text{START\_TIME} \gets \lceil \text{  NEGOTIATE\_INTERVAL} + len(\text{CHUNKS}) * \text{PULSE\_INTERVAL } \rceil  $
    \State \text{ }

    \State \text{\underline{\textbf{2. Determining Initial Sender and Receiver.}}}
    \State \text{\textbf{Both sites:}} sleep until \text{START\_TIME}
    \State \text{\textbf{Both sites:}} consume resources until pool is exhausted
    \If{``Site A'' is able consume over $>$ 50\% of pool}
      \State $\text{SENDER} \gets \text{``Site A''}$
      \State $\text{RECEIVER} \gets \text{``Site B''}$
    \Else
      \State $\text{SENDER} \gets \text{``Site B''}$
      \State $\text{RECEIVER} \gets \text{``Site A''}$
    \EndIf
    \State \textbf{Sender:} consumes 100\% of pool resources
    \State \textbf{Receiver:} releases all pool resources
    \State \textbf{Both sites:} sleep until $\text{START\_TIME} + \text{NEGOTIATE\_INTERVAL}$
    \State \text{ }
    \State \text{\underline{\textbf{3. Sending Data.}}}
      \For{$i \gets 0..len(\text{CHUNKS})$}
        \State Sleep until $\text{START\_TIME} + \text{NEGOTIATE\_INTERVAL} + i * \text{PULSE\_INTERVAL}$
          \If{SELF == SENDER}
        \State SEND\_INT $\gets binaryToDecimal(\text{CHUNK}[i])$
        \State Consume all unheld resources in pool
        \State Release SEND\_INT resources in the pool
    \ElsIf{SELF == RECEIVER}
        \State Sleep for $0.5 * \text{PULSE\_INTERVAL}$
        \State Consume all unheld resources in pool
        \State $\text{RECV\_INT} \gets \text{number of consumed resources}$
        \State Release all held pool resources
        \State $\text{RECV\_STR } \gets decimalToBinary(\text{RECV\_INT})$
        \State $\text{RECV\_MSG} \mathbin\Vert\text{=}\text{ RECV\_STR}$
      \EndIf
    \EndFor
    \State Release all held pool resources
    \State \text{ }

  \end{algorithmic}
  \caption{General algorithm for a \TechniqueName{} attack.}
  \label{algo:actual-algo}
\end{algorithm}

We present a generic protocol for conducting a \TechniqueName{} attack
over \LBU{} resource pools in all browsers. This protocol is presented as
Algorithm \ref{algo:actual-algo}, and provides a generic way a site can
use a \LBU{} resource pool to track users.

\SubPoint{Protocol Inputs}
The algorithm takes several inputs. First, the algorithm takes which resource pool will be
exploited to conduct the attack, which also determines the size of the pool.
Attackers can precompute the largest pool available for each browser.
The size of the resource pool
(\IE{} the number of resources in the pool available) is stored as \texttt{POOL\_SIZE}.

The second input is the message being sent over the channel,
which is a binary string of arbitrary length. The binary string to be transmitted is stored as \texttt{MSG}.

Third, the algorithm takes two time intervals, stored
as \texttt{NEGOTIATE\_INTERVAL} and \texttt{PULSE\_INTERVAL}. These intervals could be fixed across all attack methods,
and trade faster transmission time (smaller values) against higher reliability (lower values).

\SubPoint{Step One: Setup}
To begin, the sender splits \texttt{MSG} into
\texttt{PKT\_SIZE} sized chunks, yielding a vector of bit-strings each of size
\texttt{PKT\_SIZE}, and the receiver constructs an empty buffer,
\texttt{RECV\_MSG}, to accumulate the received message into one packet at a time.
The sending and receiving sites must choose the same time to start communication: the shared \texttt{START\_TIME} is set to the next integer ECMAScript epoch time (in seconds) that is greater than a multiple of the full negotiation and message transmission time. 

\SubPoint{Step Two: Determining Initial Sender and Receiver}
Both parties synchronize by sleeping until the \texttt{START\_TIME}, and then determine which site will be the initial \Sender{}
and which the \Receiver{}. This negotiation is needed because, neither
site initially knows what other colluding site(s) may be open and available
to communicate with, and thus no way of assigning roles in the protocol.

Sites determine \Sender{} and \Receiver{}
by racing to exhaust the resource pool. The site that is able to consume more
than~$50\%$ of resource in the pool assigns itself the role of initial \Sender{};
the site that is prevented from requesting the~$50\%+1$ resource assigns
itself as the initial \Receiver{}.

The \Receiver{}
then releases the resources it holds, and the \Sender{} keeps consuming
resources until the pool is exhausted.

\SubPoint{Step Three: Sending Data}
The third step of the protocol is where passing data across context (\IE{} site)
boundaries occurs. The \Sender{} and the \Receiver{} participate in this step
of the protocol as follows.

The \Sender{} manipulates the state of the resource pool as follows for
each $c$ in their \texttt{CHUNKS} vector (recall that \texttt{CHUNKS} is
a vector of binary strings, each of length \texttt{PKT\_SIZE}). For each
$c$, the \Sender{} first interprets the binary as positive
integer representation (\EG{} $0010010$ becomes 18, etc), which is stored
as \texttt{SEND\_INT}. The \Sender{} then releases $\texttt{SEND\_INT} + 1$ resources
from the pool and waits for a fixed period, \texttt{PULSE\_INTERVAL}, to ensure that the the \Receiver{}
has had time to read from the channel. Once the \Sender{} has finished sending their message, the \Sender{} releases
all resources in the pool and proceeds to the next step in the protocol.
Otherwise, the \Sender{} consumes all resources in the pool and repeats
the current stage in the protocol to send the next $c$ value.

Simultaneously, the \Receiver{} begins this stage of the protocol by waiting for \texttt{PULSE\_INTERVAL}/2.
Once that time has elapsed, the \Receiver{}
tries to consume as many resources as possible, which will match the
\texttt{SEND\_INT} number of resources released by the \Sender{}, and stores
this value (minus 1) as \texttt{RECV\_INT}\footnote{Recall that, by construction, the \Sender{} is not able
to obtain more than $2^{PKT\_SIZE}$ resources, and so the \Receiver{} can
be certain that values greater than the limit, or equal to zero, are not data the \Sender{} is attempting
to transmit. If so, the \Receiver{} will exit the protocol. } Next, the \Receiver{}
encodes \texttt{RECV\_INT} in the inverse
manner the \Sender{} used (\EG{} 18 becomes $0010010$), and concatenates the
result onto the \Receiver{'s} \texttt{RECV\_MSG}. The \Receiver{} then releases
all resources it holds, waits for the end of the pulse, and repeats the above process.

\SubPoint{Step Four: Exchanging Roles}
Finally, if desired, the two parties can exchange roles to pass data in the
\Receiver{} to \Sender{}. This is trivially accomplished by each party assuming
the opposite role, and continuing again from step 3. Otherwise, if there is
no more data to transmit, both parties can abort the protocol. Note that
the protocol itself does not provide a mechanism for the parties to indicate
whether they wish to continue or end the protocol, though parties could easily
signal such through the contents of the messages being passed.

\section{Evaluation in Popular Browsers}
\label{sec:eval}
In the previous section we presented a generic algorithm for turning \LBU{}
resource pools into cross-context communication channels, which in turn
can be used to cookie-sync and track users across the \Web{}. In this section
we demonstrate three examples of such \LBU{} resource pools in popular browsers,
and measure how exploitable and practical they are for cross-site tracking.

Specifically, we show that practical forms of \TechniqueName{} attacks can
be carried out in popular browsers. We implemented three examples of
\TechniqueName{} attacks, using the
WebSockets,
Server Sent Events (SSE),
and Web Workers APIs.
Before this work, all Chromium browsers were vulnerable to the WebSockets and SSE attacks,
Firefox was vulnerable to the Web Sockets and Web Workers attacks, Tor Browser was vulnerable
to the WebSockets attack, and Safari was vulnerable to the SSE attack.

We assess the practicality of each of implemented \TechniqueName{} attack through
four measurements: \textbf{Availability}, the size of the relevant resource pool, and
the kind of context-linking possible, \textbf{bandwidth}, or how long it takes to send a~35-bit
identifier through the channel, \textbf{consistency}, or how often the identifier is sent correctly,
and \textbf{background noise}, or how often sites on the \Web{} use resources
in each resource pool.

% First, we discuss
% the \textbf{availability} of the attack
% in each browser engine, meaning we present the size of the resource pool
% in each browser engine and what kinds of context-linking
% are possible using that resource pool in each browser engine (\EG{} cross-site
% communication, cross-profile communication, etc.). Second, we measure the
% \textbf{bandwidth} of the communication channel, by measuring how long our
% algorithm takes to transmit a unique identifier
% (encoded as a bit string) across contexts with high fidelity. Third, because browser resource pools do not always precisely enforce the \LBU{} constraint, we measure how often a 35-bit string can be accurately transmitted across contexts in our best implementation.
% Fourth, we measure the incidence of \textbf{background noise} with respect to the particular \TechniqueName{} attack,
% by conducting a representative crawl of the \Web{}, and measuring how frequently
% we observe how often the relevant Web API is used on the \Web{} currently (and
% so, how much background noise an attacker might have to contend with when conducting
% the attack).

Finally, for all measurements of Chromium browsers, we configured each browser
to enable all site-as-privacy-boundary features enabled (\IE{} we enabled
all browser features designed to allow communication across sites or site-partitions).
Specifically, we disabled third-party cookies, and enabled all ``Network Isolation Key''
features to enforce cache and network-state partitioning (see Section \ref{sec:definitions:contrasts:stateful}
for more information on these ``hardened'' Chromium configurations see). We note the exception
here was Brave, which partitions DOM storage and network state by default.

% A brief summary of the browsers and their underlying engines, and hence
% the applicability of the below measurements, is presented
% in Table \ref{table:browser-engines}.

\subsection{Attack Availability}

\begin{table*}[t]
  \begin{center}
    \small
        \begin{tabular}{lllrrr}  
            \toprule
        \bf Browser     & \bf Engine    & \bf Version & \bf WebSockets & \bf Web Workers & \bf Server-Sent Events \bigstrut[t]\\
            \midrule
        Brave       & Chromium  & 1.44.101  & $*$ 255     & -      & 1,350 \bigstrut[t]\\
        Chrome      & Chromium  & 105.0.5195.125          & 255        & -      & 1,350 \\
        Edge        & Chromium  & 106.0.1370.42          & 255        & -      & 1,350 \\ 
        Firefox     & Gecko     & 105.0.1          & $\dagger$ 200       & 512    & - \\
        Safari      & WebKit    & 15.2          & -          & -      & $*$ 6 \\
        Tor Browser & Gecko     & 11.5.2        & 200        & -      & - \\
        \bottomrule
    \end{tabular}
  \end{center}
  \caption{\textbf{Attack Availability}: Size of each resource pool used to 
  conduct each instance of a \TechniqueName{} attack. ``-'' denotes that
  the browser was not vulnerable. ``$*$'' indicates that the vulnerability was fixed before
  this work was submitted. ``$\dagger$'' denotes that the resource-pool can be exploited to
  conduct cross-profile attacks (in addition to cross-site).}
  \label{table:availability}
\end{table*}

\SubPoint{Methodology}
We first checked the availability (and so, exploitability) of each example
\TechniqueName{} attack by experimenting with
browsers and examining the source code of each browser engine to identify \LBU{} resource pools in those browsers. 

Specifically, we considered Web APIs that might hypothetically represent a finite pool of resources (network connections, threads, etc.). We then used the developer console in each browser to manually test whether each Web API would leak and whether that resource pool could be predictably exhausted. We conducted these tests as follows:

\begin{enumerate}
    \item We opened a new tab,. visited a blank page, and checked
    if instantiating the candidate Web API repeatedly in a loop in the developer console caused errors after a finite and predictable number of calls.
    \item Once that pool was exhausted, we then opened a second tab to a different site.  Did we find that no more resources were available under the second site?
    \item If we released a resource under the first tab, could a single resource now be consumed without error under the second tab?
    \item Were we able to find logic in the corresponding browser engine code that was imposing this resource pool limit?
\end{enumerate}
If the answer to all four of these questions was yes, then we concluded this Web API was vulnerable to the pool party attack for the tested browser. We thus proceeded to implement attacks against each vulnerable browser based on the
algorithm presented in Section~\ref{sec:algo}, implemented in \JS{}\footnote{\SourceLink{}}.
We then examined whether we could use the relevant resource pool to create a covert-channel and communicate across
site boundaries, across profile boundaries, or both. Importantly, we tested whether
the attack technique can be used to
communicate between a standard-browsing profile, and a ``private
browsing mode'' profile\footnote{This feature goes by different names in
different browsers, but generically refers to the ability to run the browser
in a way where stored values only last the lifetime of the browsing session.}.

\SubPoint{Results}
Our \emph{availability} measurements yielded several significant findings, summarized in Table~\ref{table:availability}.

First, we were able to identify exploitable \LBU{} resource pools in
all major browsers, which we were successfully able to exploit through
\TechniqueName{} attacks (though Safari and Brave both fixed some vulnerabilities during the
``responsible disclosure'' process).
As noted, the resource pools targeted in each browser engine differ. We were able
to use the relatively large WebSockets connection pool in Chromium-
and Gecko-based browsers to conduct \TechniqueName{} attacks.  Safari's WebSockets implementation was not exploitable,
since WebKit does not restrict how many WebSocket connections can be opened simultaneously. 
Safari's implementation of the SSE API, though, was previously exploitable before they fixed it.
(Gecko's implementation of the SSE API \emph{was not} exploitable).

Firefox alone was vulnerable to the Web Workers form of the attack (a surprising finding given
that Tor Browser uses the same Gecko engine).

Second, we found that Gecko-based browsers (\IE{} Firefox and Tor Browser)
were vulnerable to \TechniqueName{} attacks in a way more concerning than other browser engines. 
While \TechniqueName{} attacks can be used for cross-site tracking in
all browsers, \textbf{in Gecko-based browsers \TechniqueName{} attacks can be used to track
users \emph{across profiles}}. 
Significantly, this means that, in Gecko-based browsers, sites can
conduct \TechniqueName{} attacks between private browsing sessions and standard browsing sessions. 
More concretely, a site running in a private browsing window can collude with
a site running in a standard browsing window, and identify both sessions as belonging to
the same person. This is particularly concerning since it violates the core promise of
a private browsing session; that behaviors conducted using a private browsing are ``ephemeral'',
and cannot be linked other accounts or behaviors a user maintains. Additionally, this
vulnerability undermines the work and research that has been done to strengthen private browsing modes
in browsers (for example, \cite{burzstein2010analysis, lerner2013verifying, gao2014private}).

\subsection{Attack Bandwidth}
\label{sec:eval:bandwidth}

\begin{table}[t]
  \begin{center}
    \small
    \begin{tabular}{llrrrr}
      \toprule
\bf Browser     & \bf Method     & \bf Setup & \bf Send & \bf Total & \bf Success \bigstrut[t]\\
      \midrule
        Brave       & SSE        &  3.0                &         5.0     &      8.0            &  100\%    \bigstrut[t]\\
        Chrome      & SSE        &  2.0                &         5.0     &      7.0            &  100\%    \\
        Edge        & SSE        &  2.0                &         5.0     &      7.0            &  100\%    \\
      \midrule
        Chrome      & WS  &   0.1               &         0.5     &       0.6            &  100\%    \\
        Edge        & WS  &   0.1               &         0.5     &       0.6            &  100\%    \\
        Firefox     & WS  &  2.0                &         5.0     &      7.0            &   71\%    \\
        Tor Browser & WS  &  2.0                &         5.0     &      7.0            &   73\%    \\
      \midrule
        Firefox     & WW     &  1.5            &         7.5     &      9.0            &   95\%    \\
      \bottomrule
    \end{tabular}
  \end{center}
  \caption{\textbf{Attack Bandwidth}: How long it took to transmit a 35-bit string using Server-Sent Events (SSE), WebSockets (WS), or WebWorkers (WW). Times are reported in seconds; all values are reported over 100 runs.}
  \label{table:bandwidth}
\end{table}

\SubPoint{Methodology}
We measured the bandwidth of each attack by measuring how long
each implemented \TechniqueName{} attack took to transmit a 35-bit
string across the site (or in the case of Firefox's WebSockets implementation, profile) boundaries.

We selected a 35-bit string for two reasons.
First, because it is over 33-bits,  or what is needed to uniquely identify the
approximately 7.9 billion people on the planet, and two, because it
aligns cleanly with the 5-bit packet size used in WebSocket and Web Worker experiments.

We conduct each measurement as follows.
First, we manually open two tabs on a browser to two pages on two different sites
we controlled. Each page includes an
implementation of the relevant \TechniqueName{} attack
(Websockets and SSE in Chromium-based browsers, WebSockets and Web Workers
in Gecko-based browsers, and SSE in Safari), implemented through \JS{} included in the page.
We then experimentally varied the negotiation time and pulse time until
we found the minimum times necessary to ensure that messages were passed
accurately with a high success rate. We then conduct this measurement 100 times,
using a clean browser profile for each measurement, and report the average.

\SubPoint{Results}
We report the results of our bandwidth measurements in Table~\ref{table:bandwidth}.
Times are reported in seconds, and the transmission success rate (discussed in the next
subsection) is reported as a percentage.

We find that our example \TechniqueName{} attacks are practical.
Even the slowest forms of the attack complete in under ten seconds (far below the
average page dwell time of slightly under a minute\cite{liu2010understanding}).
Each attack could further be carried out between pages that are
left open for a moderate amount of time, either because they get lost in a
browser users ever-growing collection of tabs, or because the site is intended
to stay open for a long time
(\EG{} sites that function and email clients, instant messaging applications, video
streaming sites, etc). Our example \TechniqueName{} attacks are fast enough that
the could be conducted multiple times during an average page view (again, assuming an
average page dwell time of slightly under one minute), as a simple error handling
technique to account for noisy channels.

The ``Setup'' column in Table~\ref{table:bandwidth} corresponds to the
``Determining Initial Sender and Receiver''
section of Algorithm~\ref{algo:actual-algo}; the ``Send'' column
measures the ``Sending Data'' steps.

% Second, there is an unsurprising connection between the size of the relevant
% resource pool, and the bandwidth of the covert-channel that can be created over
% the resource pool. Sites can covertly-communicate through larger resource pools (such as
% Chromium's WebSocket implementation) more quickly.

% NOTE: Removing this paragraph because I actually saw some unexplained differences in the minimum messaging times.
% Third, we observe very little performance difference between different browsers using the same
% browser engine, at least within the three Chromium browsers included in this measurement.
% We note that manual tests with the Tor Browser support the same
% conclusion between Gecko based browsers, once the overhead of the Tor network is accounted
% for. Since we were unable to include the Tor Browser in our automated
% measurements though, we mention the performance similarly
% between Firefox and Tor Browser as an anecdotal observation, and not as a
% high confidence finding.

\subsection{Attack Consistency}
\SubPoint{Methodology}
We also evaluated how consistently each \TechniqueName{} attack example completed successfully,
in the absence of other sites running on the browser. This measurement provides an upper bound
on how practical the attack could be, as having other sites running in parallel in the browser
will in some cases further reduce the success rate.

We measured the consistency of each attack using the same methodology described in
Section~\ref{sec:eval:bandwidth}. We again ran the attack~100 times, on two different pages
in a single instance of the browser, each time in a clean profile. We then report the percentage
of times the~35-bit string was received correctly.

\SubPoint{Results}
The results of our consistency measurement is also reported in Table~\ref{table:bandwidth}.
We find that most forms of the attack are either perfectly consistent (\IE{} all 100 evaluations
executed correctly), or consistent enough to be practical (\IE{} the Web Worker attack on Firefox). The WebSocket attack was less consistently successful in
the Gecko browsers. We investigated the root cause and found that the pool size was not consistently enforced;
Firefox occasionally allowed additional sockets to be created, resulting in a corrupted message.

\subsection{Attack Background Noise}
\label{sec:eval:noise}

\begin{table}[t]
  \begin{center}
    \small
    \begin{tabular}{lrrr}
      \toprule
                           & \bf \% page  & \bf \% of URLs & \bf \% of URLS \bigstrut[t]\\
        \bf Web API            & \bf Loads    & \bf Desktop  & \bf Mobile   \\
      \midrule
        Web Worker         &  12.34\% &  12.29\%   & 11.9\% \bigstrut[t]    \\
        WebSocket          &  9.55\%  &   4.33\%   & 3.72\%     \\
        Server-Sent Events &  0.79\%  &   0.8\%    & 0.06\%     \\
      \bottomrule
    \end{tabular}
  \end{center}
  \caption{\textbf{Attack background noise}: Web API metrics reported by the Chrome Platform Status service, as of \ChromePlatformDate{}. Numbers reflect the \% of page loads and \% of URLs observed across all channels and platforms.}
  \label{table:stability}
\end{table}

\SubPoint{Methodology}
Finally, we evaluated how noisy the communication channels used in our demonstrative \TechniqueName{}
attacks are in practice. We build on the intuition that resource pools that are infrequently used
by sites ``in the wild'' for being purposes ) are easier to convert into practical side-channels.
Put differently, if sites are already consuming and releasing resources in a resource pool 
for benign purposes, then other sites intending to use it as a covert communication channel
have to contend with more noise and uncertainty, and thus communication will be more difficult.

We estimate an upper bound on the presence of background noise per tab
that could interfere with 
our \TechniqueName{} attacks by using HTML \& JavaScript usage metrics reported by
the Chrome Platform Status website \footnote{https://chromestatus.com/metrics/feature/popularity} 
to look at how often \Web{} sites
use WebSocket, Web Worker, and/or SSE capabilities.

We note that we initially measured background use on the \Web{} through an automated, \Web{}-scale crawl,
using browsers instrumented to count how often the Web Workers, WebSockets and Server-Sent Events
APIs were used. However, we abandoned this approach on realizing that an automated crawl
would potentially \emph{under report} background noise, since in some cases sites would only use
these ``advanced'' browser capabilities on user interaction. This realization lead us to
instead look for measurements of browsers under real-world use, and thus to Chrome telemetry.

\SubPoint{Results}
We report how often the relevant browser APIs are used during real-world browser
use (as reported by ``Chrome Platform Status'') in Table~\ref{table:stability}.
Reported numbers are of \ChromePlatformDate{}.
As noted, no browser feature is used on most websites; one reported feature, SSE,
is used on less than~1\% of websites, and less than~1\% of page loads.
Put differently, the vast majority of sites do not use any
of these browser features, meaning that in the common case sites could use the
resource pool without any interference from other pages. 

\SubPoint{Notes and qualifications}
In practice, sites colluding in a \TechniqueName{} attack
would need to contend with the union of all open sites accessing resources from
the relevant resource pool. Browser with large numbers of tabs open are therefore
more likely to present interference to the two attacking tabs. The numbers resulting
from our methodology are a lower bound on how noisy the given resource
pool would be, and attackers might need to implement ways of communicating over noisy channels.

Additionally, we note that resources used by a site during the duration of a \TechniqueName{} attack
will not effect correctness, only bandwidth.
Held resources merely reduce the total limit on a resource pool, but our algorithm can still proceed to send messages.
Only resources that go from unconsumed to consumed by a site (or vise versa) during the attack
will affect correctness.

Further, we note that the faster an attack completes, the less susceptible the attack is
to errors introduced by background noise. This is for two reasons. First, attacks that
finish quickly are less likely to be interrupted with benign background resource use (simply
because there is less opportunities for background resource use). And second, attacks that
finish quickly can engage in simple error correcting techniques to account for possible 
background noise (for example, conducting the attack multiple times and taking the majority
result).

Nonetheless, some percentage of sites are likely to be calling the Web APIs in a more dynamic
manner, so it's useful to understand how often these features are used in the wild.

\section{Discussion}
\label{sec:discussion}

\subsection{Implementation Challenges}
We have shown that, a \LBU{} resource pool whose resources can be consumed and released
by scripts in web pages are sufficient to allow cross-site communication.
A few additional anomalies arose, however, in the implementation of
our algorithm that would be necessary to consider
for the development of a robust message-passing script.

\subsection{Drifting Resource Pool Limit} Because the global limit on a resource pool is not a central feature of web browsers, it
is possible that developers may overlook certain behaviors of the resource pool. For
example, we found that the Firefox global WebSocket pool limit was not strictly fixed.
While our script continuously created and destroyed WebSockets to manipulate the
number of unheld WebSocket vacancies in the pool, we observed that occasionally
the total number of WebSockets that could be created increased. That is: while
the initial limit on WebSockets was~512, after a few cycles of the algorithm,
the total number of resources that could be held was observed to be~513. That
number continued to increase over time. We attribute this behavior to a likely
race condition that meant the limit on the total number of allow WebSockets
was not consistently enforced, but occasionally a WebSocket slipped through and
was not counted.

If the limit changed during a message cycle, then at least
one value passed from sender to receiver in our implementation
would be incorrect. That would result in
an incorrect 35-bit message being recorded by the receiver.
To minimize the effects of this anomaly, we ensured that the sender repeatedly attempted to 
consume more resources than the expected limit, so that even if the limit silently
increased during one cycle, the the algorithm would correctly pass the message in
subsequent cycles.

\SubPoint{Delayed feedback} A second anomaly we observed in the browser-\JS{}
implementation of our algorithm was an inconsistent delay in feedback from the resource pool
when the pool limit was hit. If a site tries to consume an $n$-th resource when only $n-1$
resources were available, the attack script must receive an indication that the limit has been
reached for the algorithm to succeed.

For example, to consume a resource in the WebSocket pool, it is necessary for the script
to call \texttt{new WebSocket(...)}. In all cases, whether or not a limit
on the WebSocket pool has already been reached, the call returns a WebSocket object.
To determine whether the limit has been reached, it's necessary to find out whether
the WebSocket object is in a valid state or not. The state can be ascertained by using
the \texttt{onerror} callback property on the WebSocket object; however this callback
is not fired after a consistent time
interval. Instead the time interval varied by as much as tens of milliseconds in
some cases. Therefore, it is necessary to introduce a delay before checking for
the presence of an error state.

In order to avoid introducing too much delay when attempting to consume $n$ resources,
we first create all $n$ resource objects in a tight loop, and then wait for success or failure
for all of the resources in parallel.

\subsection{Additional Attack Vectors}
\label{sec:discussion:additional}

In this work we demonstrated that \TechniqueName{} attacks are possible and
practical in all popular browsers, by exploiting the \LBU{} implementations
of WebSockets, Web Workers, and Server-Sent Events. However, there are many other
\TechniqueName{} attack opportunities in current browsers. This section details
some additional APIs and browser capabilities that can be converted into
covert-channels though
\TechniqueName{} attacks. We do not intend this to be a comprehensive list; we
expect that there are many more \TechniqueName{} attack vectors in browsers.
We identified the below browser capabilities and APIs as likely exploitable
by \TechniqueName{} attacks based on their
implementations in Gecko, WebKit and / or Chromium, though
we did not build attacks to test the exploit-ability of all listed APIs.
% source leads
% us to suspect that some-or-all of these capabilities could be leveraged for
% \TechniqueName{} attacks.

\SubPoint{Chromium} Chromium's DNS resolver has a global, unpartitioned
limit of 64 simultaneous requests, which could be exploited by an attacker
that could control the response time of DNS queries.
Second, when Chromium browsers are configured to use an HTTP proxy, they
impose a global limit of 32 simultaneous network requests. Third, several APIs
in Chromium are thin-wrappers around OS-provided systems, and maintain a single
global handle to the system process, effectively creating
\LBU{} pool of size one (\EG{} the Web Speech API).

\SubPoint{Gecko} Browsers built on Gecko have many of the same additional
attack vectors as Chromium based browsers; Gecko has a global
limit on the number of DNS requests that can be in the air at the same time,
and a lower limit on the number of requests that can be open when
using a HTTP(S) proxy.

\SubPoint{WebKit} We identified far fewer additional \LBU{} resource pools in
WebKit than in other browser engines. We did not identify additional
attack vectors in Safari (the most popular WebKit browser) beyond SSEs. However,
other, less popular browsers also use WebKit, and some of these
introduce additional \LBU{} resource pools. For example, the
GTK-based version of WebKit\footnote{\url{https://trac.webkit.org/wiki/WebKitGTK}}
uses a DNS resolver with a limit of 8 requests at a time,
and a pre-fetch cache with a limit of 64 hosts.

\subsection{Defenses and Constraints}
Defending against \TechniqueName{} attacks in browsers is difficult since,
at root, systems will always have limited resources, and thus some underlying
\LBU{} pool that can be exploited by a sufficiently motivated party. Currently
browser resource limitations are mostly explicit and intentional, but even
if browsers removed such limits (\EG{} limits on WebSocket connections),
the underlying system would necessarily have a global limit, either explicitly
(\EG{} OS imposed limitations on open network sockets) or implicitly
(\EG{} systems have a finite amount of memory, and so unavoidably can only
maintain a finite number of network sockets).

However, even if \TechniqueName{} attacks will fundamentally always be possible,
browsers can still take steps to limit how practical such attacks
might be.

One approach is to lift browser-imposed limits on resource
pools where applicable, and require attackers to contend with much larger
system-maintained resource pools. This approach would make attacks much more obvious to
the user, who would notice their system slowing down or other applications on
the system impacted while the attack was being carried out. Such detectability
might deter attackers, at least in the common case. Relying on the OS or system
level limits would also make attacks more difficult to carry out. For example,
the system network connection pool will be much ``noisier'' than the browser's
WebSocket pool, making it much more difficult to use the resource pool as
a reliable covert channel.

Second, browsers could take the opposite approach, and instead of dramatically
widening the size of resource pools, browsers could maintain existing
resource caps but partition resource pools the same way
browsers increasingly partition DOM storage and network state (see
Section~\ref{sec:definitions:contrasts}). If resource pools were partitioned by
site, a site would not learn anything by exhausting its resource pool; the
resource pools for other sites would be unaffected. A determined attacker
could regain the ability to conduct a \TechniqueName{} attack by controlling
a large number of sites, and using them to collectively drain the resource pool.
Nevertheless, partitioning resource pools by site would make \TechniqueName{}
attacks significantly more difficult for an attacker to carry out.

Third, browsers could combine these two approaches, and simultaneously remove global
limits on the size of resource pools, but limit the number of resources
each site or context and use. Such a hybrid approach would achieve some of the
benefits of both of the above approaches.

\subsection{Applicability to Mobile}
As part of this work, we checked whether mobile versions of each of
browser were also vulnerable to \TechniqueName{} attacks. To do so,
we checked that the availability of the WebSockets and SSE attacks on mobile
versions of each browser matched the availability of each attack on the
desktop version. We found that the availability of each attack was the same;
attacks that worked on the desktop version of a browser also worked on the
browser's mobile browser, and attacks that did not work on the desktop version
also did not work on mobile.

We did not measure the bandwidth or stability of the identified \TechniqueName{} attack on mobile, both because i. of limited resources, and ii. it being somewhat more difficult to conduct automated measurements on mobile browsers (\EG{} most automation tools target desktop versions of browsers, or rely on imperfect simulations of mobile environments). Assessing the practicality of \TechniqueName{} attacks on mobile browsers is an important area for future work.

\SubPoint{Identifying Resource Pools}
The vulnerabilities discussed in this work were identified through a combination
of domain expertise, source code review, and manually interaction with each API in each browser.

We began by generating a list of candidate APIs, based on our experience in both
browser and Wbb-app development. Our list of candidate APIs focused on features
that i. needed to be handled in parallel (\EG{} threads, I/O operations),
ii. use limited system resources (\EG{} file handles) or iii. are exclusive by
nature (\EG{} only one voice can speak at a time when using the Web Speech API).

Next, once we had our set of candidate APIs, we examined the source code for API's implementation in each browser engine. We looked for explicit limits on resources, both in the source code and surrounding comments. Often (though not always), limits were relatively easy to find, since they were encoded in constants or runtime flags.

Finally, we manually evaluated whether we could use these browser limits to conduct pool-party attacks by constructing two different test pages on different sites and seeing if we could detect on one site when the relevant resources were being exhausted by the other site.

As noted, this process is entirely manual; it is likely-to-certain that there are more vulnerable resource pools in browsers. Developing a system for systematically or automating the detection of vulnerable resource pools would be a valuable area for future work.

\section{Related Work}
\label{sec:related}

\SubPoint{Online tracking through feature misuse}
Our work exists alongside a large body of work on ways browser features
can be (mis)used by online trackers, to track their users in ways
unintended by the browser vendor.

The largest volume of work in this area is on browser fingerprinting. 
We highlight significant work in the area, specifically those that identified new fingerprinting vulnerabilities in browsers.
Mowery \EtAl{}~\cite{mowery2012pixel} famously demonstrated that differences in how
browsers executed drawing (\IE{} canvas and WebGL) operations could be used to identify individuals,
and Acar \EtAl{}~\cite{acar2013fpdetective} showed that differences in what fonts users
have installed could be similarly misused. Englehardt \EtAl{}~\cite{englehardt2016online},
as part of a project to measure privacy violations
on the 1m most popular websites, show the WebAudio and WebRTC APIs
could be used to track users. Olejnik \EtAl{}~\cite{olejnik2015leaking} showed the Battery
Status API could be used for fingerprinting, and Olejnik and Janc~\cite{olejnik2017stealing} demonstrated
the Ambient Light API could be similarly misused. Zhang \EtAl{}~\cite{zhang2019sensorid}
found that in mobile browsers, websites can use unpermissioned access to  motion
sensors to identify users, and Starov and Nikiforakis~\cite{starov2017xhound}
demonstrated that what browser extensions the user has installed can make users more
identifiable. Eckersley~\cite{eckersley2010unique} documented that a screen resolution
and display size were practical fingerprinting vectors, and Nikiforakis \EtAl{}~\cite{nikiforakis2013cookieless} showed that other 
display details contributed to identifiability. Laperdrix \EtAl{}~\cite{laperdrix2016beauty}
found that, ironically, the presence of a content blocker could help fingerprinters distinguish users. Iqbal
\EtAl{}~\cite{iqbal2021fingerprinting} identified dditional browser APIs that
fingerprinting methods misuse (\EG{} proximity sensor APIs, media capabilities, the Presentation API)
by examining the \JS{} source code of known fingerprinting scripts and identifying the
additional APIs those scripts abuse.

Distinct from browser fingerprinting, researchers have found other ways of
misusing browser features to construct u user identifiers. Solomos \EtAl{}~\cite{favicon2021}
transformed the browser's ``favicon'' cache into a persistent tracking mechanism,
Janc \EtAl{}~\cite{janc2020information} showed that Safari's ``Intelligent Tracking
Prevention''\footnote{\url{https://webkit.org/blog/7675/intelligent-tracking-prevention/}} features could be abused to re-identify users, and Syverson and Traudt~\cite{syverson2018hsts} showed how the
browsers ``HTTP Strict-Transport-Security'' system could be re-purposed to construct and
assign unique identifiers. 

A parallel body of work attempts to prevent browser fingerprinting.
Nikiforakis \EtAl{}~\cite{nikiforakis2015privaricator} found browsers could resist
fingerprinting by manipulating the values common Web APIs return. Laperdrix \EtAl{}~\cite{laperdrix2017fprandom} extended this approach by introducing small amounts of noise into the Web Audio
and Canvas APIs, changes large enough to cause fingerprinters to misidentify users, but
small enough that benign uses of features would not be impacted. Snyder \EtAl{}~\cite{snyder2017most}
suggested disabling Web APIs whose cost to the users (including additional
identifiability) was higher than the benefit to them (in terms of desirable page behaviors),
based on prior work finding that most browser APIs are rarely used at all~\cite{snyder2016browser}.
Smith \EtAl{}~\cite{smith2021sugarcoat} suggest a strategy for rewriting malicious code to
prevent fingerprinting scripts from accessing the underlying, identifying values. Other works
aim to prevent attackers from abusing features for tracking purposes by removing
the entropy added by OS or hardware differences. Wu \EtAl{}~\cite{wu2019rendered} presents
a method for removing hardware-induced differences in WebGL operations, and though not
targeting fingerprinting attacks, Andrysco \EtAl{}~\cite{andrysco2018towards} proposed
a similar ``improve privacy by making dissimilar systems execute similarly'' approach
for floating-point based channels in browsers.

%\vspace{-0.4cm}

\SubPoint{Attacks on browser partitioning and sandboxing}
Our work also builds on, and exists along side, a large body of work documenting
ways browser partitioning efforts can be circumvented, whether those
partitions are enforced directly by the application, through OS-based process
isolation, or otherwise. Again, the breath of
work in this area makes a comprehensive discussion here impossible, so we discuss
papers that are particularly significant, novel and/or recent.

Using timing methods to circumvent browser partitioning (in this case the same-origin-policy)
dates back at least as far as 2000, when Felten \EtAl{}~\cite{felten2000timing} presented
a way sites could determine what other sites the user had visited by probing the
HTTP cache and measuring timing differences. Bortz \EtAl{}~\cite{bortz2007exposing} published
similar foundational work on how sites could exploit timing differences in how, and how
quickly, cross-site requests and resources were loaded to to learn about users' state on other sites.
Since then, researchers have demonstrated many ways sites can circumvent browser imposed restrictions on what sites can learn
about user behavior on other sites, and how sites can communicate with each other.

Schwartz \EtAl{}~\cite{schwarz2017fantastic}
document ways sites can create high resolution timers, with descriptions
for how such timers can form covert channels across application boundaries. Smith
\EtAl{}~\cite{smith2018browser} show that sites can circumvent browsers attempts to partition
a users browsing history by exploiting cache state and side effects in painting behaviors.
Kohlbrenner and Shacham~\cite{kohlbrenner2017effectiveness} presented a way of creating a
covert channel across site boundaries by exploiting floating-point related timing channels
in how browsers render SVGs. Gruss \EtAl{}~\cite{gruss2016rowhammer} extended the Rowhammer~\cite{kim2014flipping} attack,
previously used to leak information across OS process boundaries, to be exploitable be
through site-included \JS{} code, to violate browser imposed process isolation.
Jin \EtAl{}~\cite{jin2022siteisolation} show how security-focused
protections like isolating sites in their own OS processes can be exploited to learn what
sites the user is, or has recently, visited. Lipp \EtAl{}~\cite{lipp2017practical} demonstrated
how sites could puncture site isolation protections and infer what the user was typing
on a different site (or different application) by observing timing patterns in \JS{} execution,
caused by the OS responding to key presses issued to other contexts. Vila
\EtAl{}~\cite{vila2017loophole} presented a related attack, where a site could transform contention
in the browser's main event loop (distinct from the event loop presented to an executing \JS{}
context) into a cross-context side channel. van Goethem \EtAl{}~\cite{van2015clock} showed
how other browser capabilities like service workers and the (now deprecated) application cache
can be transformed into timer-based covert channels as well.
As part of a larger project of creating an automated
system for detecting cross-site information leaks, Knittel \EtAl{}\cite{knittel2021xsinator}
identified how many other browser features that had side effects that could be detected
across site boundaries.

The cross-profile tracking attack against Gecko-based browsers described in this work build
on other cross-profile attacks such as van Goethem and Joosen~\cite{van2017one}, which
found that application efforts to isolate ``incognito'' browsing sessions from
standard browsing sessions could be circumvented by using contention for disk and
memory resources as a covert channel.  Oren \EtAl{}~\cite{oren2015spy} showed that contention
in the CPU cache could be exploited by unprivileged, malicious \JS{} code to learn
what sites a user was visiting in an ``incognito'' mode session.
with \JS{} running on other sites, in other processes, or applications outside the browser.
Gruss \EtAl{}~\cite{gruss2015practical} present a similar attack, though instead targeting
timing channels stemming from memory deduplication.

Finally, Asankah~\cite{asankah2020ephemeral} defines a cross-site identification
technique they call ``ephemeral fingerprinting'', where different sites observe infrequent global events to those infrequent events to reidentify a user across contexts.

\vspace{-0.2cm}

\section{Conclusions}
\label{sec:conclusions}

In this work we define a new category of practical privacy attack in popular \Web{} browsers
we call \TechniqueName{} attacks. \TechniqueName{} attacks allow sites to break out
of the ``contextual sandboxes'' that browsers try to enforce, and so allow sites
to circumvent privacy protections in even the most aggressively privacy-focused browsers.
More alarming still, we find that \TechniqueName{} techniques can be used to track
users beyond cross-site tracking (specifically, that in Gecko-based browsers \TechniqueName{}
attacks can track users \emph{across profiles}).

While some attacks in this category have been known to be \emph{theoretically} possible,
this work demonstrates that such attacks are \emph{practical}, and must be dealt with as
a real-world threat to the \Web{} users' privacy. Further, we show that \TechniqueName{}
attacks can be carried out using a wider range of browser capabilities than previously
documented, further emphasizing the severity of the risk to user privacy.

% After much research on \Web{} privacy from academia, research and activists, we are
% encouraged and excited that many browser vendors seem to be entering a virtuous
% cycle, where browsers compete to see who can ship new features to more aggressively
% protect user's privacy on the Web. We hope that this work aids browser vendors in these
% laudable efforts.

Web privacy has moved in two very different directions over the last two decades.
\emph{Privacy attacks} have moved in a dispiriting direction, with
privacy violations becoming common place. This disappointing trend is reinforced by a combination of conflicting incentives from (some) browser vendors, backwards compatibility concerns, and user-harming financial incentives. \emph{Privacy defenses} in browsers, though, have been recently moving in an encouraging direction. This is due to (in part) a combination of regulatory pressure, increasing user awareness, the tireless efforts of privacy-focused researchers and developers, and a virtuous competition between (some) browsers to own the ``most private browser'' title. We hope that this work helps the latter, to the determent of the former.
\section{Acknowledgements}
We'd like to thank Deian Stefan and Michael Smith from University of California,
San Diego for contributing additional examples of \LBU{} resource pools in
browsers. We'd also like to thank Rainer Böhme from University of Innsbruck for refining and improving this work.
\balance
\newpage
\bibliographystyle{plain}
\bibliography{paper}
\end{document}